\documentstyle[12pt]{article}

\author{\sc
       P.M. Lavrov$^{a)}$\thanks{E-mail: lavrov@tspu.edu.ru},\,\,\,\,L.A. Malysheva$^{a)}$,
       \,\,P.Yu. Moshin$^{a),b)}$
       \\
       \\
       \normalsize\it
       $^{a)}$Tomsk State Pedagogical University, Tomsk 634041, Russia
       \\
       \normalsize\it
       $^{b)}$Instituto de F{\'{\i}}sica, Universidade de S\~{a}o Paulo,\\
       \normalsize\it
       Caixa Postal 66318-CEP, 05315-970 S\~{a}o Paulo, S.P., Brazil}
\title{\LARGE\bf
       Generalized Superfield Lagrangian Quantization}
\date{}
\begin{document}
\maketitle

\begin{quotation}
\noindent
We consider an extension of the gauge-fixing procedure
in the framework of the Lagrangian superfield BRST and BRST-antiBRST
quantization schemes for arbitrary gauge theories, taking into account the
possible ambiguity in the choice of the superfield antibracket. We show
that this ambiguity is fixed by the algebraic properties of the antibracket
and the form of the BRST and antiBRST transformations, realized in terms of
superspace translations. The Ward identities related to the generalized
gauge-fixing procedure are obtained.
\end{quotation}

\section{Introduction}

Much attention has been paid recently to setting up a superspace framework
for gauge theory quantization (see, e.g.,  \cite{attention,LMR,L}). The
inspiration for these studies is provided by the interpretation of the BRST
\cite{BRST} and antiBRST \cite{antiBRST} symmetries in terms of superspace
translations, first realized for theories with a gauge group \cite{BT}.

In the papers \cite{LMR,L}, a superfield Lagrangian formalism for arbitrary
gauge theories was proposed. Namely, \cite{LMR} provides a superfield
description of the BV formalism \cite{BV}, based on the BRST symmetry, while
\cite{L} is a superfield form of the Sp(2) covariant scheme
\cite{BLT}, based on the extended BRST symmetry (including BRST and antiBRST
transformations).

 The superspace framework \cite{LMR,L} is based on a choice of gauge-fixing
 that allows one to combine the variables of the quantum theory into
 supervariables. These supervariables are applied to introduce superspace
 counterparts of the basic ingredients of the original schemes
 \cite{BV,BLT}, i.e. the antibracket and its generating operator. Along
 with these ingredients, a crucial role in the superfield formalism belongs
 to a set of operators that generate the transformations of supervariables
 related to superspace translations. The (extended) BRST symmetry is
 realized as the transformations generated by these new operators,
 accompanied by the transformations induced by the antibracket.

 Because the supervariables contain an extended set of field components, in
 comparison with the set of field-antifield variables \cite{BV,BLT}, they
 can be used to construct different algebraic structures similar to the
 antibracket. This raises a natural question of a possible ambiguity in a
 superfield description of the quantization rules \cite{BV,BLT} along the
 lines of \cite{LMR,L}. This question is related to generalizations of the
 superfield approach \cite{LMR,L}, respecting the principle of the
 gauge-independence of the S-matrix.

 In reference with such generalizations, it is relevant to take into
 account an extension of the original BV and Sp(2) covariant schemes
 proposed in \cite{BBD,3pl,mod3pl}. Namely, it was demonstrated that a
 possible way of extending the BV formalism and the Sp(2) covariant
 approach, in a manner respecting the gauge-independence of the S-matrix,
 is to impose the gauge with the help of a special generating
 equation for the action of gauge-fixing.

 In this paper, we generalize the superfield approach \cite{LMR,L} along the
 lines of \cite{BBD,3pl,mod3pl}, taking into account the possible ambiguity
 in the choice of the superfield antibracket, respecting the superalgebra
 satisfied by the operators generating the antibracket and superspace
 translations \cite{LMR,L}. We demonstrate that the form of the antibracket
 is fixed by the superalgebra \cite{LMR,L}, with allowance for the
 requirement of (extended) BRST symmetry, encoding the gauge-independence
 of the S-matrix. The compatible form of the antibracket is limited to its
 representations found in \cite{LMR,L}.

 The resulting quantization formalism provides an extension of the
 gauge-fixing procedures used  in the original BV and Sp(2) covariant
 schemes \cite{BV,BLT}, which is exemplified by a generalized form of the
 Ward identities. At the same time, the present generalization of the
 approach \cite{LMR,L} can be regarded as a superfield description of the
 studies \cite{BBD,mod3pl}.

 The paper is organized as follows. In Sections 2, 3, we present the
 generalization of the superfield formalism \cite{LMR,L} and discuss
 the corresponding transformations of (extended) BRST symmetry.
 In Section 4, we use these transformations to derive the Ward identities
 and prove the gauge independence of the S-matrix. In Section 5,
 we discuss the relation of the generalized superfield formalism to the
 quantization schemes \cite{LMR,L,BV,BLT,BBD,mod3pl}.

 We apply the conventions used in \cite{LMR,L}. Integration over
 supervariables is assumed as integration over their components.
 Derivatives with respect to (super)fields are taken from the right (unless
 otherwise specified), while those with respect to (super)antifields and
 supersources are taken from the left. At the classical level, we consider
 an arbitrary gauge theory, implying the well-known structure of the space
 of fields $\phi^A$, $\varepsilon(\phi^A)=\varepsilon_A$, given by \cite{BV}.

\section{Generalized Superfield BRST Quantization}
\setcounter{equation}{0}

In this section, we shall generalize the procedure of superfield BRST
quantization \cite{LMR}. We begin by introducing a superspace
$(x^\mu,\theta)$, where $x^\mu$ are space-time coordinates, and $\theta$ is
an anticommuting coordinate. Let $\Phi^A(\theta)$ be a set of superfields
$\Phi^A(\theta)$, associated with a set of the corresponding
super-antifields $\Phi^*_A(\theta)$, with the Grassmann parities
$\varepsilon(\Phi^A)\equiv\varepsilon_A,\,
\varepsilon(\Phi^*_A)=\varepsilon_A+1$, and satisfying the boundary
condition
\begin{eqnarray}
\label{BondSQ}
 \Phi^A(\theta)|_{\theta =0}=\phi^A.
\end{eqnarray}
Define the vacuum functional $Z$ as the following path integral:
\begin{eqnarray}
\label{ZSQ}
Z=\int
 d\Phi\;d\Phi^*\rho(\Phi^*)
\exp\bigg\{\frac{i}{\hbar}\bigg[W(\Phi,\Phi^*) +
X(\Phi,\Phi^*) +S_0(\Phi,\Phi^*)\bigg]\bigg\},
\end{eqnarray}
where $W=W(\Phi,\Phi^*)$ is the quantum action determined by the generating
equation
\begin{eqnarray}
\label{GEq2SQ}
\bar{\Delta}\exp\bigg\{\frac{i}{\hbar}W\bigg\}=0
\end{eqnarray}
and $X=X(\Phi,\Phi^*)$ is a Bosonic gauge-fixing functional satisfying
the equation
\begin{eqnarray}
\label{GEqX1SQ}
\tilde{\Delta}\exp\bigg\{\frac{i}{\hbar}X\bigg\}=0.
\end{eqnarray}

In (\ref{ZSQ}), we have used a functional $\rho(\Phi^*)$, which defines
the weight of integration over the super-antifields $\Phi^*_A(\theta)$ and has
the form of a functional $\delta$-function:
\begin{eqnarray}
\label{WFuncSQ}
 \rho(\Phi^*)&=&\delta\bigg(\int d\theta\,\Phi^*(\theta)\bigg).
\end{eqnarray}
Besides, we have introduces the functional
\begin{eqnarray}
\label{p*p}
 S_0(\Phi,\Phi^*)=\Phi^*\Phi&=&\int
d\theta\,\Phi^*_A(\theta)\Phi^A(\theta).
\end{eqnarray}

Notice that integration over the anticommuting coordinate $\theta$
is defined as follows:
\begin{eqnarray}
\nonumber
 \int d\theta=0,\;\;\int d\theta\;\theta=1,
\end{eqnarray}
which implies that any function $f(\theta)$
\[
 f(\theta) = f_0+\theta f_1
\]
admits the representation
\begin{eqnarray}
\nonumber
 f(\theta) = \int d\theta^{'}\delta(\theta^{'}- \theta)f(\theta^{'}),\;\;
 \delta(\theta^{'}-\theta)=\theta^{'}-\theta
\end{eqnarray}
and satisfies the identity
\[
 \int d\theta\,\frac{\partial f(\theta)}{\partial\theta}=0,
\]
with the consequent property of integration by parts
\[
 \int d\theta\,\frac{\partial f(\theta)}{\partial\theta}g(\theta)=
 -(-1)^{\varepsilon(f)}\int d\theta\,f(\theta)\frac{\partial g(\theta)}
 {\partial\theta},
\]
where derivatives with respect to $\theta$ are taken from the left.

In (\ref{GEq2SQ}), (\ref{GEqX1SQ}), we have used the notations
$\bar{\Delta}$, $\tilde{\Delta}$ for Fermionic operators of the form
\begin{eqnarray}
\label{Deltas_nSQ}
\bar{\Delta}=\Delta + \frac{i}{\hbar}V,\;\;
\tilde{\Delta}=\Delta - \frac{i}{\hbar}U,
\end{eqnarray}
where $U$, $V$ are first-order differential operators, assumed to have
the properties of nilpotency and anticommutativity
\begin{eqnarray}
\label{NilSQ}
 U^2=0,\;\;V^2=0,\;\;UV+VU=0,
\end{eqnarray}
whereas $\Delta$ is a second-order differential operator subject to the algebraic
conditions
\begin{eqnarray}
\label{AlgSQ}
 {\Delta}^2=0,\;\;\Delta U+U\Delta=0,\;\;\Delta V+V\Delta=0.
\end{eqnarray}

With allowance for (\ref{Deltas_nSQ}), equations (\ref{GEq2SQ}), (\ref{GEqX1SQ})
can be recast in the equivalent form
\begin{eqnarray}
\label{GEq1SQ}
\frac{1}{2}(W,W)+VW&=&i\hbar\Delta W,\\
\label{GEqXSQ}
\frac{1}{2}(X,X)-UX&=&i\hbar\Delta X,
\end{eqnarray}
where $(\,\,,\,)$ denotes an antibracket defined by the action of the
operator $\Delta$ on the product of arbitrary functionals $F$, $G$
\begin{eqnarray}
\label{antibr1}
 (F,G)=(-1)^{\varepsilon(F)} \Delta(FG)- (-1)^{\varepsilon(F)}
 (\Delta F)G-F(\Delta G).
\end{eqnarray}
The antibracket possesses the properties
\begin{eqnarray}
\label{antibprop1}
 &&\varepsilon((F,G))=\varepsilon(F)+\varepsilon(G)+1,\nonumber\\
 &&(F,G)=-(-1)^{(\varepsilon (F)+1)(\varepsilon (G)+1)}(G,F),\nonumber\\
 &&D(F,G)=(DF,G)-(F,DG)(-1)^{\varepsilon(F)},\\
 \nonumber\\
\label{fgh1}
 &&(F,GH)=(F,G)H+(F,H)G(-1)^{\varepsilon(G)\varepsilon(H)},\\
 \nonumber\\
\label{Jacobi1}
 &&((F,G),H)(-1)^{(\varepsilon (F)+1)(\varepsilon (H)+1)}
 +{\rm cycle}\,(F,G,H)\equiv 0,
\end{eqnarray}
where $D=(\Delta,U,V)$.

Notice that (\ref{antibprop1}) is a consequence of (\ref{AlgSQ}) and
(\ref{antibr1}). The property (\ref{fgh1}) follows from the fact
that $\Delta$ in (\ref{antibr1}) is a second-order differential operator,
while the Jacobi identity (\ref{Jacobi1}) follows from
(\ref{AlgSQ})--(\ref{fgh1}).

With allowance for (\ref{NilSQ}), (\ref{AlgSQ}), (\ref{antibprop1}),
the operators $\bar{\Delta}$, $\tilde{\Delta}$, defined by
(\ref{Deltas_nSQ}), possess the properties
\begin{eqnarray*}
 \bar{\Delta}^2=0,\,\,\,\,
 \tilde{\Delta}^2=0,\,\,\,\,
 \bar{\Delta}\tilde{\Delta}+\tilde{\Delta}\bar{\Delta}=0
\end{eqnarray*}
and
\begin{eqnarray*}
 \bar{\Delta}(F,G)&=&(\bar{\Delta}F,G)
 -(F,\bar{\Delta}G)(-1)^{\varepsilon(F)},\nonumber\\
 \tilde{\Delta}(F,G)&=&(\tilde{\Delta}F,G)
 -(F,\bar{\Delta}G(-1)^{\varepsilon(F)}.
\end{eqnarray*}

To find a manifest realization of equations (\ref{GEq2SQ}),
(\ref{GEqX1SQ}), or, equivalently, (\ref{GEq1SQ}), (\ref{GEqXSQ}), we shall
assume that $U$ and $V$ are identified with the operators which generate
the transformations of supervariables induced by the translations
$\theta\to\theta+\mu$ with respect to the anticommuting coordinate:
\begin{eqnarray*}
\delta\Phi^A(\theta)&=&\mu\frac{\partial \Phi^A(\theta)}
{\partial \theta}=\mu U\Phi^A(\theta),\\
\delta{\Phi}_A^*(\theta)&=&\mu\frac{\partial{\Phi}_A^*(\theta)}
{\partial \theta}=\mu V{\Phi}_A^*(\theta).
\end{eqnarray*}
The generators $U$ and $V$ are first-order differential operators, having the
form of $\theta$-local functionals
\begin{eqnarray}
 U&=&-\int d\theta \frac{\partial\Phi^A(\theta)}
 {\partial\theta}\frac{\delta_l}{\delta\Phi^A(\theta)},\nonumber\\
\label{VSQ}
 V&=&-\int d\theta\frac{\partial\Phi^*_A(\theta)}{\partial\theta}
 \frac{\delta}{\delta\Phi^*_A(\theta)},
\end{eqnarray}
where
\begin{eqnarray*}
 &&\frac{\delta_l\Phi^A(\theta)}{\delta\Phi^B(\theta^{'})}
 =(-1)^{\varepsilon_A} \delta(\theta^{'}- \theta)\delta^A_B
 =(-1)^{\varepsilon_A}\frac{\delta\Phi^A(\theta)}{\delta\Phi^B(\theta^{'})},\\
 &&\frac{\delta\Phi^*_A(\theta)}{\delta\Phi^*_B(\theta^{'})}
 =(-1)^{\varepsilon_A +1} \delta(\theta^{'}- \theta)\delta^B_A.
\end{eqnarray*}
As a consequence, the manifest form (\ref{VSQ}) implies the required algebraic
properties (\ref{NilSQ}).

Let us now construct an operator $\Delta$ satisfying the conditions
(\ref{AlgSQ}). For this purpose, we consider the class of Fermionic
differential operators of second-order, such that the entire dependence
on the components of $\Phi^A(\theta)$ and ${\Phi}_A^*(\theta)$ enters
through the functional derivatives
\begin{eqnarray}
\label{deriv1}
 \frac{\delta}{\delta\Phi^A(\theta)},\,\,
 \frac{\delta}{\delta{\Phi}_A^*(\theta)}.
\end{eqnarray}
Consider the subclass of $\theta$-local functionals with the integrands
constructed from various combinations of $\theta$,
$\frac{\partial}{\partial\theta}$, and the derivatives (\ref{deriv1}),
where summation over the indices $A$ is assumed. Notice that the analysis
of such expressions is simplified with the help of integration by parts and
the use of the anticommutator
$\{\theta,\frac{\partial}{\partial\theta}\}=1$. The specified subclass
contains only one linearly independent operator satisfying the conditions
(\ref{AlgSQ}) imposed on $\Delta$. This operator can be chosen in the form
\begin{eqnarray}
\label{DeltaSQ}
\Delta=-\int d\theta(-1)^{\varepsilon_A}\frac{\delta_l}
 {\delta\Phi^A(\theta)}
 \frac{\partial}{\partial\theta}\frac{\delta}
 {\delta\Phi^*_A(\theta)},
\end{eqnarray}
which generates the corresponding antibracket
\begin{eqnarray}
\label{ABSQ}
 (F,G)=\int d\theta\bigg\{\frac{\delta F}{\delta\Phi^A(\theta)}
 \frac{\partial}{\partial\theta}\frac{\delta
 G}{\delta\Phi^*_A(\theta)}(-1)^{\varepsilon_A+1}
 -(-1)^{(\varepsilon(F)+1)(\varepsilon(G)+1)}(F\leftrightarrow
 G)\bigg\}.
\end{eqnarray}

From the set of generating equations (\ref{GEq1SQ}), (\ref{GEqXSQ}), given
in terms of the manifestly specified ingredients $U$, $V$, $\Delta$,
$(\,\,,\,)$, it follows that the integrand in the vacuum functional
(\ref{ZSQ}) is invariant under the transformations of global supersymmetry
\begin{eqnarray}
\label{BRSTSQ}
 &&\delta\Phi^A(\theta)=\mu U\Phi^A(\theta)+(\Phi^A(\theta), X - W)\mu,
\nonumber\\
 &&\delta\Phi^*_A(\theta)=\mu V\Phi^*_A(\theta)+(\Phi^*_A(\theta), X - W)\mu,
\end{eqnarray}
with an anticommuting parameter $\mu$.

Let us analyse the variation of the integrand in the vacuum functional
(\ref{ZSQ}) with respect to the transformations (\ref{BRSTSQ}). To examine
the change of the exponential in (\ref{ZSQ}), we notice that
\begin{eqnarray}
\label{generl1}
 \delta(W+X+S_0)=\mu\bigg((W,W)-(X,X)+(U+V)(W+X)+(S_0,W-X)\bigg).
\end{eqnarray}
 To examine the change of the integration measure in (\ref{ZSQ}), we
 observe that the weight functional $\rho({\Phi}^*)$ (\ref{WFuncSQ}) is
 invariant under the transformations (\ref{BRSTSQ}),
 $\delta\rho({\Phi}^*)=0$, while the corresponding Jacobian $J$ has the form
\begin{eqnarray}
\label{J1}
 J = \exp (2\mu\Delta W - 2\mu\Delta X).
\end{eqnarray}

 Denote by $I$ the integrand in (\ref{ZSQ}). Then, with allowance for
 (\ref{GEq1SQ}), (\ref{GEqXSQ}), (\ref{generl1}) and (\ref{J1}), its
 variation $\delta I$ with respect to the transformations (\ref{BRSTSQ}) is
 given by
\begin{eqnarray}
\label{deltaI1}
 \delta I=i\hbar^{-1}\mu I\bigg((U-V)(W-X)+(S_0,W-X)\bigg).
\end{eqnarray}
Taking into account the explicit form of the operators $U$, $V$
and the antibracket $(\,\,,\,)$, given by (\ref{VSQ}), (\ref{ABSQ}),
we observe the identity
\[
 (S_0,F)=(V-U)F,
\]
where $F$ is an arbitrary functional. Therefore, according to
(\ref{deltaI1}), we find $\delta I=0$, and the integrand in (\ref{ZSQ})
is actually left invariant by the transformations (\ref{BRSTSQ}).
The transformations (\ref{BRSTSQ}) are the transformations of BRST symmetry
in the superfield formalism with the gauge fixed by a solution of the
generating equation (\ref{GEqXSQ}).

\section{Generalized Superfield BRST--antiBRST Quantization}
\setcounter{equation}{0}

Consider now a generalization of the superfield BRST-antiBRST quantization
\cite{L}. To this end, we shall introduce a superspace with coordinates
$(x^\mu,\theta^a)$, where $x^\mu$ are space-time coordinates, and
$\theta^a\;(a=1,2)$ are anticommuting coordinates. Let us consider
superfields $\Phi^A{(\theta)}$, $\varepsilon (\Phi^A)=\varepsilon_A$,
subject to the boundary condition (\ref{BondSQ}), and supersources
$\bar{\Phi}_A{(\theta)}$ possessing the same Grassmann parity,
$\varepsilon(\bar{\Phi}_A)=\varepsilon_A$. Let us define the vacuum
functional $Z$ as the following path integral:
\begin{eqnarray}
\label{ZExSQ}
Z=\int d\Phi d\bar{\Phi}\rho (\bar{\Phi})\exp \Bigg\{
\frac{i}{\hbar}\bigg[W(\Phi,\bar{\Phi})+ X(\Phi,\bar{\Phi})
+ S_0(\Phi,\bar{\Phi})\bigg]\Bigg\},
\end{eqnarray}
were $W=W(\Phi,\bar{\Phi})$ is the quantum action determined by the generating
equations
\begin{eqnarray}
\label{GEqExSQ}
\bar{\Delta}^a\exp\bigg\{\frac{i}{\hbar}W\bigg\}=0
\end{eqnarray}
and $X=X(\Phi,\bar{\Phi})$ is a Bosonic gauge-fixing functional satisfying
the equations
\begin{eqnarray}
\label{GEqXExSQ}
 \tilde{\Delta}^a\exp\bigg\{\frac{i}{\hbar}X\bigg\}=0.
\end{eqnarray}
In (\ref{ZExSQ}), we have introduced functionals $\rho(\bar{\Phi})$
and $S_0(\bar{\Phi},\Phi)$, given by
\begin{eqnarray}
\label{WegFExSQ}
 \rho(\bar{\Phi})&=&\delta\left(\int d^2\theta\,\bar{\Phi}(\theta)\right),\\
\label{BilFExSQ}
 S_0(\bar{\Phi},\Phi)&=&\bar{\Phi}\Phi=
 \int d^2\theta\,\bar{\Phi}_A(\theta)\,\Phi^A(\theta).
\end{eqnarray}

 Notice that integration over the anticommuting coordinates $\theta^a$
 is defined as follows:
\[
 \int d^2\theta=0,\;\;\int d^2\theta\;\theta^a=0,\;\;\int d^2\theta\;
 \theta_a\theta^b=\delta^b_a,
\]
 where raising and lowering the Sp(2) indices is carried out by the rule
 $\theta^a=\varepsilon^{ab}\theta_b$, $\theta_a=\varepsilon_{ab}\theta^b$,
 with $\varepsilon^{ab}$ being a constant antisymmetric tensor,
 $\varepsilon^{12}=1$. The above definitions imply that any function $f(\theta)$
\[
 f(\theta)=f_0+\theta^a f_a+\theta^2 f_3,\,\,\,
 \theta^2\equiv\frac{1}{2}\theta_a\theta^a
\]
 admits the representation
\[
 f(\theta)=\int d^2\,\theta'\,\delta(\theta'-\theta)f(\theta'),\,\,\,
 \delta(\theta'-\theta)=(\theta'-\theta)^2
\]
 and satisfies the identity
\[
 \int d^2\theta\;\frac{\partial f(\theta)}{\partial \theta^a}=0,
\]
 with the consequent property of integration by parts
\[
\int d^2\theta\;\frac{\partial f(\theta)}{\partial \theta^a}g(\theta)=
-\int d^2 \theta (-1)^{\varepsilon(f)}f(\theta)\frac{\partial g(\theta)}
{\partial\theta^a}\,,
\]
where derivatives with respect to $\theta^a$ are taken from the left.

In (\ref{GEqExSQ}), (\ref{GEqXExSQ}), we have used the notations
$\bar{\Delta}^a$, $\tilde{\Delta}^a$ for doublets of Fermionic
operators of the form
\begin{eqnarray}
\label{Deltas_nExSQ}
 \bar{\Delta}^a=\Delta^a + \frac{i}{\hbar}V^a,\;\;
 \tilde{\Delta}^a=\Delta^a - \frac{i}{\hbar}U^a,
\end{eqnarray}
where $U^a$, $V^a$ are first-order differential operators,
assumed to have the properties of nilpotency and anticommutativity
\begin{eqnarray}
\label{algUV}
 U^{\{a}U^{b\}}=0,\;\;V^{\{a}V^{b\}}=0,\;\;V^aU^b+U^bV^a=0,
\end{eqnarray}
while $\Delta^a$ is a doublet of second-order differential operators
subject to
\begin{eqnarray}
 &&\Delta^{\{a}\Delta^{b\}}=0,\nonumber\\
\label{algDelta}
 &&\Delta^{\{a}V^{b\}}+V^{\{a}\Delta^{b\}}=0,\nonumber\\
 &&\Delta^{\{a}U^{b\}}+U^{\{a}\Delta^{b\}}=0,
\end{eqnarray}
with symmetrization over Sp(2) indices in (\ref{algUV}), (\ref{algDelta})
introduced according to $A^{\{a}B^{b\}}=A^aB^b+A^bB^a$.

With allowance for (\ref{Deltas_nExSQ}), the generating equations
(\ref{GEqExSQ}), (\ref{GEqXExSQ}) can be recast in the equivalent form
\begin{eqnarray}
\label{GEq1ExSQ}
 \frac{1}{2}(W,W)^a+V^aW&=&i\hbar\Delta^aW,\\
\label{GEqX1ExSQ}
 \frac{1}{2}(X,X)^a - U^aX&=&i\hbar\Delta^aX,
\end{eqnarray}
where $(\,\,,\,)^a$ denotes an extended antibracket defined by the
action of the operator doublet $\Delta^a$ on the product of
arbitrary functionals $F$, $G$
\begin{eqnarray}
\label{antibr}
 (F,G)^a=(-1)^{\varepsilon(F)} \Delta^a(FG)- (-1)^{\varepsilon(F)}
 (\Delta^a F)G-F(\Delta^a G).
\end{eqnarray}
The extended antibracket possesses the properties\footnote{These
relations are analogous to the properties (\ref{antibprop1})--(\ref{Jacobi1})
of the previous section.}
\begin{eqnarray}
\label{antibprop}
 &&\varepsilon((F,G)^a)=\varepsilon(F)+\varepsilon(G)+1,\nonumber\\
 &&(F,G)^a=-(-1)^{(\varepsilon (F)+1)(\varepsilon (G)+1)}(G,F)^a,\nonumber\\
 &&D^{\{a}(F,G)^{b\}}=(D^{\{a}F,G)^{b\}}-(F,D^{\{a}G)^{b\}}
 (-1)^{\varepsilon(F)},\\
 \nonumber\\
 &&(F,GH)^a=(F,G)^aH+(F,H)^aG(-1)^{\varepsilon(G)\varepsilon(H)},\nonumber\\
 &&((F,G)^{\{a},H)^{b\}}(-1)^{(\varepsilon (F)+1)(\varepsilon (H)+1)}
 +{\rm cycle}\,(F,G,H)\equiv 0,\nonumber
\end{eqnarray}
where $D^a=(\Delta^a,U^a,V^a)$.

With allowance for (\ref{algUV}), (\ref{algDelta}), (\ref{antibprop}),
the operators $\bar{\Delta}^a$, $\tilde{\Delta}^a$, defined by
(\ref{Deltas_nExSQ}), possess the properties
\begin{eqnarray*}
\label{AlgExSQ}
 \bar{\Delta}^{\{a}\bar{\Delta}^{b\}}=0,\,\,\,\,
 \tilde{\Delta}^{\{a}\tilde{\Delta}^{b\}}=0,\,\,\,\,
 \bar{\Delta}^{\{a}\tilde{\Delta}^{b\}} +
 \tilde{\Delta}^{\{a}\bar{\Delta}^{b\}}=0,
\end{eqnarray*}
and
\begin{eqnarray*}
\label{bartilde}
 \bar{\Delta}^{\{a}(F,G)^{b\}}&=&(\bar{\Delta}^{\{a}F,G)^{b\}}
 -(F,\bar{\Delta}^{\{a}G)^{b\}}(-1)^{\varepsilon(F)},\nonumber\\
 \tilde{\Delta}^{\{a}(F,G)^{b\}}&=&(\tilde{\Delta}^{\{a}F,G)^{b\}}
 -(F,\bar{\Delta}^{\{a}G)^{b\}}(-1)^{\varepsilon(F)}.
\end{eqnarray*}

To find a manifest realization of equations
(\ref{GEqExSQ}), (\ref{GEqXExSQ}), or, equivalently, (\ref{GEq1ExSQ}),
(\ref{GEqX1ExSQ}), we shall assume that $U^a$ and $V^a$ are identified
with the operators which generate the transformations of supervariables
induced by the translations $\theta^a\to\theta^a+\mu^a$ with respect to
the anticommuting coordinates:
\begin{eqnarray*}
\delta\Phi^A(\theta)&=&\mu_a\frac{\partial \Phi^A(\theta)}
{\partial \theta_a}=\mu_aU^a\Phi^A(\theta),\\
\delta\bar{\Phi}_A(\theta)&=&\mu_a\frac{\partial\bar{\Phi}_A(\theta)}
{\partial \theta_a}=\mu_aV^a\bar{\Phi}_A(\theta).
\end{eqnarray*}

The generators $U^a$ and $V^a$ are first-order differential operators,
having the form of $\theta$-local functionals
\begin{eqnarray}
 \label{U&V}
 U^a&=&\int d^2\theta\frac {\partial\Phi^A(\theta)}{\partial\theta_a}
 \frac {\delta_l}{\delta\Phi^A(\theta)},\nonumber\\
 V^a&=&\int d^2\theta\frac {\partial\bar{\Phi}_A(\theta)}{\partial
 \theta_a}\frac {\delta}{\delta\bar{\Phi}_A(\theta)},
\end{eqnarray}
where
\begin{eqnarray*}
\nonumber
&&\frac{\delta_l\Phi^A(\theta)}{\delta\Phi^B(\theta^{'})}
=\delta(\theta^{'}-\theta)\delta^A_B
=\frac{\delta\Phi^A(\theta)}{\delta\Phi^B(\theta^{'})},\\
\nonumber
&&\frac{\delta\bar{\Phi}_A(\theta)}{\delta\bar{\Phi}_B(\theta^{'})}
=\delta(\theta^{'}-\theta)\delta^B_A.
\end{eqnarray*}
As a consequence, the manifest form (\ref{U&V}) implies the required algebraic
properties (\ref{algUV}).

An explicit form of the extended operator $\Delta^a$ with the properties
(\ref{algDelta}) is {\it not unique}. Consider the class of Fermionic
second-order differential operators such that the entire dependence on the
components of $\Phi^A(\theta)$ and $\bar{\Phi}_A(\theta)$ enters through
the functional derivatives
\[
 \frac{\delta}{\delta\Phi^A(\theta)},\,\,
 \frac{\delta}{\delta\bar{\Phi}_A(\theta)}.
\]
 In the specified class, there exist only two linearly independent Sp(2)
 doublets having the form of $\theta$-local
 functionals\footnote{We consider integrands constructed from different
 combinations of $\theta^a$, $\frac{\partial}{\partial\theta^a}$,
 and the functional derivatives $\frac{\delta}{\delta\Phi^A(\theta)}$,
 $\frac{\delta}{\delta\bar{\Phi}_A(\theta)}$, assuming summation over
 the indices $A$. Notice that the analysis of such expressions is simplified
 with the help of integration by parts and the use of the anticommutator
 $\{\theta^a,\frac{\partial}{\partial\theta^b}\}=\delta^a_b$.}
 and possessing the algebraic properties (\ref{algDelta}) of the extended
 delta-operator. Because of an additional property of anticommutativity with
 each other, these operators span a two-dimensional linear space of operators
 possessing the properties of $\Delta^a$. The basis elements $\Delta^a_1$,
 $\Delta^a_2$ of this space can be presented in the form
\begin{eqnarray}
\label{DeltaaExSQ}
 &\!\!\!\!\!\!\!\!\!\!\!
 &\Delta^a_1=-\int d^2\theta \frac {\delta_{\it l}}{\delta\Phi^A(\theta)}
 \frac {\partial}{\partial \theta_a}\frac
 {\delta}{\delta\bar{\Phi}_A(\theta)}\,,\\
 &\!\!\!\!\!\!\!\!\!\!\!\!\!\!\!\!\!\!\!
\label{DeltaaExSQ2}
 &\Delta^a_2=\int d^2\theta \frac {\delta_{\it l}}{\delta\Phi^A(\theta)}
 \frac{\partial^2}{\partial\theta^2}\left(\theta^a
 \frac{\delta}{\delta\bar{\Phi}_A(\theta)}\right)\!,
\end{eqnarray}
 where
\[
 \frac{\partial^2}{\partial\theta^2}\equiv\frac{1}{2}\varepsilon^{ab}
 \frac{\partial}{\partial\theta^b}\frac{\partial}{\partial\theta^a}.
\]

 The extended delta-operators (\ref{DeltaaExSQ}) and (\ref{DeltaaExSQ2})
 generate the corresponding extended antibrackets
\begin{eqnarray}
\label{ABExSQ}
 &\!\!\!\!\!\!\!\!\!\!\!\!
 &(F,G)^a_1=\int d^2\theta\Bigg\{\frac{\delta F}{\delta\Phi^A(\theta)}
 \frac{\partial}{\partial\theta_a}\frac{\delta G}
 {\delta\bar{\Phi}_A(\theta)}(-1)^{\varepsilon_A+1}
 -(-1)^{(\varepsilon (F)+1)(\varepsilon (G)+1)}(F\leftrightarrow G)\Bigg\}
\end{eqnarray}
 and
\begin{eqnarray}
\label{ABExSQ2}
 &\!\!\!\!\!\!\!\!\!\!\!\!\!\!\!\!\!\!\!\!
 &(F,G)^a_2=\int d^2\theta\Bigg\{\!\!\left(\frac{\partial^2}{\partial\theta^2}
 \frac{\delta F}{\delta\Phi^A(\theta)}\right)\theta^a
 \frac{\delta G}{\delta\bar{\Phi}_A(\theta)}(-1)^{\varepsilon_A}
 -(-1)^{(\varepsilon (F)+1)(\varepsilon (G)+1)}(F\leftrightarrow G)\!\Bigg\}.
\end{eqnarray}

 We define the transformations of extended BRST symmetry as the following
 transformations of global supersymmetry:
\begin{eqnarray}
\label{BRSTExSQ}
 && \delta\Phi^A(\theta)=\mu_a U^a\Phi^A(\theta) + (\Phi^A(\theta),
 X - W)^a\mu_a,
 \nonumber\\
 &&\delta\bar{\Phi}_A(\theta)=\mu_a V^a\bar{\Phi}_A(\theta) +
 (\bar{\Phi}_A(\theta), X - W)^a\mu_a,
\end{eqnarray}
 providing a straightforward generalization of BRST transformations
 (\ref{BRSTSQ}), given in the previous section. Before the introduction of
 transformations (\ref{BRSTExSQ}), the manifest form of the extended
 antibracket (\ref{ABExSQ}) has no obvious advantage over (\ref{ABExSQ2}).
 It can be shown, nevertheless, that (\ref{ABExSQ}) is compatible with the
 given form of extended BRST symmetry, while (\ref{ABExSQ2}) is not.

 Let us examine the variation of the integrand in the vacuum functional
 (\ref{ZExSQ}) with respect to the transformations (\ref{BRSTExSQ}). To
 this end, notice that in both cases of the antibracket realization,
 (\ref{ABExSQ}), (\ref{ABExSQ2}), we have
\begin{eqnarray}
\label{generl}
\!\!\!\delta(W+X+S_0)\!=\!\mu_a\bigg((W,W)^a-(X,X)^a+(U^a+V^a)(W+X)
                +(S_0,W\!-\!X)^a\bigg),
\end{eqnarray}
 while the weight functional $\rho(\bar{\Phi})$ (\ref{WegFExSQ})
 is invariant under these transformations, $\delta\rho(\bar{\Phi})=0$,
 and the Jacobian $J$ has the form
\begin{eqnarray}
\label{J}
 J = \exp (2\mu_a\Delta^a W - 2\mu_a\Delta^a X),
\end{eqnarray}
 where the operator $\Delta^a$ corresponds to the choice of the
 antibracket.

 Denote by $I$ the integrand in the vacuum functional (\ref{ZExSQ}). Then,
 with allowance for (\ref{GEq1ExSQ}), (\ref{GEqX1ExSQ}), (\ref{generl})
 and (\ref{J}), its variation $\delta I$ with respect to (\ref{BRSTExSQ})
 is given by
\begin{eqnarray}
\label{deltaI}
 \delta I=i\hbar^{-1}\mu_a I\bigg((U^a-V^a)(W-X)+(S_0,W-X)^a\bigg).
\end{eqnarray}

 The  invariance of the integrand depends on a manifest choice of the extended
 antibracket. Thus, in the case of the antibracket (\ref{ABExSQ}) the condition
 $\delta I=0$ is satisfied in accordance with the identity
\[
 (S_0,F)^a=(V^a-U^a)F,
\]
 where $F$ is an arbitrary functional. At the same time, in the case
 of the antibracket (\ref{ABExSQ2}) we have
\[
 (S_0,F)^a\not\equiv(V^a-U^a)F,
\]
 and therefore, according to (\ref{deltaI}), the integrand in (\ref{ZExSQ})
 is not invariant under the transformations (\ref{BRSTExSQ}) without
 imposing additional restrictions on the functionals $W$ and $X$.

\section{Gauge Independence and Ward Identities}
\setcounter{equation}{0}
 In this section, we shall investigate the consequences implied by the
 transformations of (extended) BRST symmetry (\ref{BRSTSQ}),
 (\ref{BRSTExSQ}). Namely, we shall apply these transformations to obtain
 the corresponding Ward identities and prove the gauge-independence of the
 S-matrix in the framework of the generalized superfield quantization.

 The study of gauge-dependence is based on the equivalence theorem
 \cite{KT} stating that the gauge-independence of the S-matrix
 is ensured by the independence of the vacuum functional under
 small gauge variations. Let us examine the gauge-dependence of the vacuum
 functional $Z$ (\ref{ZSQ}). Note, first of all, that any admissible
 variation $\delta X$ of the gauge-fixing functional $X$ should satisfy the
 equation
\begin{eqnarray*}
\label{}
(X,\delta X) - U\delta X = i\hbar \Delta \delta X,
\end{eqnarray*}
which can be recast in the form
\begin{eqnarray}
\label{VarXSQ}
 \hat{Q}(X)\delta X=0.
\end{eqnarray}
In (\ref{VarXSQ}), we have introduced a nilpotent operator $\hat{Q}(X)$,
\begin{eqnarray*}
\label{QSQ}
 \hat{Q}(X)=\hat{\cal B}(X)-i\hbar\tilde{\Delta}, \quad
 \hat{Q}^2(X) = 0.
\end{eqnarray*}
Here, $\hat{\cal B}(X)$ stands for an operator acting by the rule
\begin{eqnarray*}
\label{}
 (X,F)\equiv\hat{\cal B}(X)F
\end{eqnarray*}
and possessing the property
\begin{eqnarray*}
\label{}
 \hat{\cal B}^2(X)=\hat{\cal B}\left(\frac{1}{2}(X,X)\right).
\end{eqnarray*}
The nilpotency of the operator $\hat{Q}(X)$ implies that any functional
of the form
\begin{eqnarray*}
\label{DeltaXSQ}
 \delta X=\hat{Q}(X)\delta \Psi,
\end{eqnarray*}
with $\delta \Psi$ being an arbitrary Fermionic functional, obeys equation
(\ref{VarXSQ}). Furthermore, by analogy with the theorems proved in
\cite{BLT}, one can establish the fact that any solution of (\ref{VarXSQ}),
vanishing when all the variables entering $\delta X$ are equal to zero, has
the form (\ref{DeltaXSQ}) with a certain Fermionic functional $\delta \Psi$.

Let $Z_X\equiv Z$ be the vacuum functional (\ref{ZSQ}) corresponding to the
gauge condition chosen as the functional $X$. In the vacuum functional
$Z_{X+\delta X}$ we shall perform the change of variables (\ref{BRSTSQ})
with $\mu=\mu(\Phi,\Phi^*)$, accompanied by an additional change,
\begin{eqnarray*}
\label{}
 \delta\Phi^A=(\Phi^A,\delta Y),\quad\delta\Phi^*_A=(\Phi^*_A,\delta
Y),\quad\varepsilon(\delta Y)=1,
\end{eqnarray*}
where $\delta Y=-i\hbar\mu(\Phi,\Phi^*)$. We have
\begin{eqnarray}
\label{ZX+SQ}
 Z_{X+\delta X}=\int d\Phi\,d\Phi^* \rho(\Phi^*)
 \exp\left\{\frac{i}{\hbar}\bigg(W+X+\delta X+\delta X_1
 +\Phi^*\Phi\bigg)\right\}.
\end{eqnarray}
In (\ref{ZX+SQ}), we have denoted
\begin{eqnarray*}
\label{}
 \delta X_1=2\bigg((X,\delta Y)-U\delta Y-i\hbar\Delta\delta
 Y\bigg)=2\hat{Q}(X)\delta Y.
\end{eqnarray*}
Let the functional $\delta Y$ be chosen in the form  (notice that
$\delta X=\hat{Q}(X)\delta \Psi$)
\begin{eqnarray*}
\label{}
 \delta Y=-\frac{1}{2}\hat{Q}(X)\delta\Psi.
\end{eqnarray*}
Then we find
\begin{eqnarray*}
\label{GIndZSQ}
 Z_{X+\delta X}=Z_X,
\end{eqnarray*}
which means that the vacuum functional and the S-matrix in the generalized
superfield BRST formalism do not depend on the gauge.

To derive the Ward identities related to the BRST transformations (\ref{BRSTSQ}),
we shall introduce a superfield generating functional of Green's functions,
$Z(\Phi^*)$, defined as the following path integral:
\begin{eqnarray}
\label{GSQ}
Z(\Phi^*)=\int
d\Phi'\;d\Phi^{*\prime}\rho(\Phi^{{*\prime}})
\exp\bigg\{\frac{i}{\hbar}\bigg[W(\Phi',\Phi^{*\prime}) +
X(\Phi',\Phi^{*\prime})+(\Phi^{{*\prime}}-\Phi^*)\Phi'\bigg]\bigg\},
\end{eqnarray}
which is identical with the vacuum functional (\ref{ZSQ})
in the case of vanishing super-antifields, $Z(0)=Z$.

In (\ref{GSQ}), we shall make a change of the integration variables in the
form of the BRST transformations (\ref{BRSTSQ}). Then, using integration by
parts in (\ref{GSQ}), we obtain the following Ward identities:
\begin{eqnarray}
\label{Wardx}
\int d\theta\,\frac{\partial\Phi^*_A(\theta)}{\partial\theta}
\left<\frac{\delta X}{\delta\Phi^{*\prime}_A(\theta)}+\Phi^{\prime A}(\theta)
\right>Z(\Phi^*)=0,
\end{eqnarray}
where $\langle F(\Phi',\Phi^{*\prime})\rangle$ stands for the expectation
value
\[
 \langle F(\Phi',\!\Phi^{*\prime})\rangle Z (\Phi^*)\!=\!\!
 \int\!\!d\Phi'd\Phi^{*\prime}\!\rho(\Phi^{{*\prime}})F(\Phi',\!\Phi^{*\prime})
 \!\exp\!\left\{\!\frac{i}{\hbar}\bigg[W(\Phi',\Phi^{*\prime})\!+\!
 X(\Phi',\!\Phi^{*\prime})\!+\!(\Phi^{{*\prime}}\!-\!\Phi^*)\Phi'\bigg]\!\right\}
\]
of an arbitrary functional $F(\Phi',\Phi^{*\prime})$. The Ward identities
(\ref{Wardx}) can be recast in the form
\begin{eqnarray}
\label{Ward}
 VZ(\Phi^*)=\frac{i}{\hbar}\int d\theta\,
 \frac{\partial\Phi^*_A(\theta)}{\partial\theta}
 \left<\frac{\delta X}{\delta\Phi^{*\prime}_A(\theta)}\right>Z(\Phi^*).
\end{eqnarray}

 We now generalize the above considerations to the case of extended BRST
 symmetry. Let us examine the gauge-dependence of the vacuum functional $Z$
 (\ref{ZExSQ}), using the antibracket (\ref{ABExSQ}) and the corresponding
 delta-operator (\ref{DeltaaExSQ}), which provide the invariance of the
 integrand (\ref{ZExSQ}) under the transformations of extended BRST
 symmetry (\ref{BRSTExSQ}).

 Notice that any admissible variation $\delta X$ of the gauge functional
 $X$ must satisfy the equation
\[
 (X,\delta X)^a - U^a\delta X = i\hbar \Delta^a \delta X,
\]
 which can be recast in the form
\begin{eqnarray}
\label{VarXExSQ}
 \hat{Q}^a(X)\delta X=0.
\end{eqnarray}
 In (\ref{VarXExSQ}), we have introduced an operator $\hat{Q}^a(X)$
 possessing the property of generalized nilpotency,
\begin{eqnarray}
\label{QaExSQ}
\hat{Q}^a(X)=\hat{\cal
 B}^a(X)-i\hbar\tilde{\Delta}^a, \quad
\hat{Q}^{\{a}(X)\hat{Q}^{b\}}(X) = 0,
\end{eqnarray}
 where $\hat{\cal B}^a(X)$ stands for an operator acting by the rule
\[
 (X,F)^a\equiv\hat{\cal B}^a(X)F
\]
 and possessing the property
\[
 \hat{\cal B}^{\{a}(X)\hat{\cal B}^{b\}}(X)=
 \hat{\cal B}^{\{a}\left(\frac{1}{2}(X,X)^{b\}}\right)\!.
\]
 With allowance for the operator $\hat{Q}^a(X)$ in (\ref{QaExSQ}),
 any functional
\begin{eqnarray}
\label{DeltaXExSQ}
 \delta X=\frac{1}{2}\varepsilon_{ab}\hat{Q}^a(X)\hat{Q}^b(X)
 \delta F,
\end{eqnarray}
 parameterized by an arbitrary Boson $\delta F$, satisfies
 (\ref{VarXExSQ}). By analogy with the theorems proved in \cite{BLT}, one
 can establish the fact that any solution of (\ref{VarXExSQ}), vanishing
 when all the variables entering the functional $\delta X$ are equal to
 zero, has the form (\ref{DeltaXExSQ}) with a certain Bosonic functional
 $\delta F$.

 Denote by $Z_X\equiv Z$ the vacuum functional (\ref{ZExSQ}) corresponding
 to the choice of gauge conditions in the form of the functional $X$. In
 the vacuum functional $Z_{X+\delta X}$ we first perform the change of
 variables (\ref{BRSTExSQ}), with $\mu_a=\mu_a(\Phi,\bar\Phi)$, and then
 the additional change of variables
\[
 \delta\Phi^A=(\Phi^A,\delta Y_a)^a,\quad
 \delta\bar{\Phi}_A=(\bar{\Phi}_A,\delta Y_a)^a,\quad\varepsilon(\delta Y_a)=1,
\]
 with $\delta Y_a=-i\hbar\mu_a(\Phi,\bar\Phi)$. We have
\[
 Z_{X+\delta X}=\int d\Phi\,d\bar\Phi \rho(\bar\Phi)
 \exp\left\{\frac{i}{\hbar}\bigg(W+X+\delta X+\delta X_1
 +\bar\Phi\Phi\bigg)\right\}\!,
\]
 where
\[
 \delta X_1=2\bigg((X,\delta Y_a)^a-U^a\delta Y_a - i\hbar\Delta^a\delta
 Y_a\bigg)=2\hat{Q}^a(X)\delta Y_a\,.
\]
 Having in mind (\ref{DeltaXExSQ}), we choose the functional
 $\delta Y_a$ in the form
\[
 \delta Y_a=-\frac{1}{4}\varepsilon_{ab}\hat{Q}^b(X)\delta F.
\]
 Then we find that $\delta X+\delta X_1=0$ and conclude that the relation
 $Z_{X+\delta X}=Z_X$ holds true. Therefore, the vacuum functional and the
 S-matrix in the generalized superfield BRST-antiBRST quantization
 do not depend on the gauge.

 To derive the Ward identities corresponding to the extended BRST symmetry
 transformations (\ref{BRSTExSQ}), we shall introduce a generating
 functional of Green's functions, $Z(\bar{\Phi})$, defined as the path
 integral
\begin{eqnarray}
\label{GSQ1}
Z(\bar{\Phi})=\int
d\Phi'\;d\bar\Phi'\,\rho(\bar\Phi')
\exp\bigg\{\frac{i}{\hbar}\bigg[W(\Phi',\bar\Phi') +
X(\Phi',\bar\Phi')+(\bar\Phi'-\bar\Phi)\Phi'\bigg]\bigg\},
\end{eqnarray}
identical with the vacuum functional (\ref{ZExSQ})
in the case of vanishing supersources, $Z(0)=Z$.

In (\ref{GSQ1}), we shall make a change of the integration variables
in the form of extended BRST transformations (\ref{BRSTExSQ}). Then,
using integration by parts in (\ref{GSQ1}), we obtain the following Ward
identities:
\begin{eqnarray*}
\label{Wardx1}
\int d^2\theta\,\frac{\partial\bar\Phi_A(\theta)}{\partial\theta^a}
\left<\frac{\delta X}{\delta\bar\Phi'_A(\theta)}+\Phi^{\prime A}(\theta)
\right>Z(\bar\Phi)=0,
\end{eqnarray*}
which can be represented in the form
\begin{eqnarray}
\label{Ward1}
V^aZ(\bar\Phi)=\frac{i}{\hbar}\int d^2\theta\,
\frac{\partial\bar\Phi_A(\theta)}{\partial\theta^a}
\left<\frac{\delta X}{\delta\bar\Phi^{\prime}_A(\theta)}\right>Z(\bar\Phi),
\end{eqnarray}
where $\langle F(\Phi',\bar\Phi')\rangle$
stands for the expectation value
\[
 \langle F(\Phi',\bar\Phi^{\prime})\rangle Z(\bar\Phi)=\!
 \int d\Phi'd\bar\Phi^{\prime}\rho(\bar\Phi^{{\prime}})F(\Phi',\bar\Phi^{\prime})
 \exp\left\{\!\frac{i}{\hbar}\bigg[W(\Phi',\bar\Phi^{\prime})+
 X(\Phi',\!\bar\Phi^{\prime})+(\bar\Phi^{{\prime}}-\bar\Phi)\Phi'\bigg]\!\right\}
\]
of an arbitrary functional $F(\Phi',\bar\Phi')$.

\section{Discussion}
\setcounter{equation}{0}

 In this paper, we have presented a generalization of the superfield Lagrangian
 formalism \cite{LMR,L} for arbitrary gauge theories on the basis
 of fixing the gauge with the help of special generating equations,
 (\ref{GEqXSQ}), (\ref{GEqX1ExSQ}).

 The present approach takes into account the possible ambiguity in the
 form of the superfield antibracket, respecting the algebraic properties
 of its generating operator (\ref{AlgSQ}), (\ref{algDelta}). In the case
 of BRST symmetry (Section 2), the antibracket is fixed (\ref{ABSQ}) by
 the properties (\ref{AlgSQ}), which leads to the BRST transformations
 (\ref{BRSTSQ}). In the case of extended BRST symmetry (Section 3), the
 antibracket is fixed (\ref{ABExSQ}) by the properties (\ref{algDelta}),
 with allowance for the requirement of extended BRST symmetry
 (\ref{BRSTExSQ}), being a straightforward generalization of
 (\ref{BRSTSQ}). In this respect, we should emphasize the remarkable
 similarity between the relations of Sections 2 and 3, with the formal
 difference in most of the cases concerning only the presence of Sp(2)
 indices.

 The formalism given in Section 2 provides an extension of the superfield
 BRST quantization \cite{LMR} in the part of gauge-fixing. Indeed, the
 operators $U$, $V$, $\Delta$ and the antibracket $(\,\,,\,)$, given by
 (\ref{VSQ}), (\ref{DeltaSQ}), (\ref{ABSQ}), coincide with the
 corresponding ingredients of the approach \cite{LMR}. At the same time,
 any functional
\[
 X(\Phi)=U \Psi(\Phi),
\]
 parameterized  by  an  arbitrary  Fermion, $\Psi=\Psi(\Phi)$, satisfies the
 generating  equation  (\ref{GEqXSQ}). This  solution  coincides  with  the
 action  of gauge-fixing used in \cite{LMR}.

 Let us analyse the component form of the superfield prescription given in
 Section 2, with the purpose of establishing its relation to other
 quantization schemes. The component form of the superfields
 $\Phi^A(\theta)$ and super-antifields $\Phi^*_A(\theta)$ is given by
\[
 \Phi^A(\theta)= \phi^A + \lambda^A\theta,\quad \Phi^*_A(\theta)=\phi^*_A -
 \theta J_A,
\]
\[
 \varepsilon(\phi^A)= \varepsilon(J_A)= \varepsilon_A, \quad
 \varepsilon(\phi^*_A)= \varepsilon(\lambda^A)= \varepsilon_A + 1.
\]
In terms of the components, the antibracket (\ref{ABSQ}) and
delta-operator (\ref{DeltaSQ})
\[
 (F,G)=\frac{\delta F}{\delta\phi^A} \frac{\delta G}{\delta \phi^*_A}
 -(-1)^{(\varepsilon(F)+1)(\varepsilon(G)+1)}(F\leftrightarrow G)\;,
\]
\begin{equation}
\label{deltabv}
 \Delta = (-1)^{\varepsilon_A}\frac{\delta_l}{\delta \phi^A} \frac
 {\delta}{\delta \phi^*_A}
\end{equation}
coincide with the corresponding ingredients of the BV formalism.
The operators $U$, $V$ in (\ref{VSQ}) are given by
\begin{eqnarray}
 U&=&-(-1)^{\varepsilon_A}\lambda^A\frac{\delta_l}{\delta \phi^A}\;,\nonumber\\
\label{V}
 V&=&-J_A \frac{\delta}{\delta \phi^*_A}.
\end{eqnarray}
Taking into account the component form of the integration measure in (\ref{ZSQ})
\begin{eqnarray}
\nonumber
 d\Phi\;d\Phi^*\,\rho(\Phi^*)=d\phi\;d\phi^*\;d\lambda\;dJ\;\delta(J),
\end{eqnarray}
and the functional $\Phi^*\Phi$ in (\ref{p*p})
\begin{eqnarray}
\nonumber
 \Phi^* \Phi = \phi^*_A \lambda^A -
 J_A\phi^A,
\end{eqnarray}
we represent the vacuum functional (\ref{GSQ}) as the following path integral:
\begin{eqnarray}
\label{GFmodBV}
 Z=\int
 d\phi\, d\phi^* d\lambda\exp\bigg\{\frac{i}{\hbar}
 \bigg[W(\phi,\phi^{*},\lambda)
 +X(\phi,\phi^{*},\lambda)+\phi^{*}_A\lambda^A \bigg]\bigg\},
\end{eqnarray}
where integration over $J_A$ has been carried out, which places
the functionals $W$, $X$ on the surface $J_A=0$.
The functionals $W=W(\phi,\phi^{*},\lambda)$, $X=X(\phi,\phi^{*},\lambda)$ satisfy
their respective generating equations (\ref{GEq1SQ}), (\ref{GEqXSQ}), restricted
to $J_A=0$. The given restriction does not affect the form of (\ref{GEqXSQ}), whereas
(\ref{GEq1SQ}) simplifies to the equation that determines the quantum action in
the BV approach,
\begin{eqnarray}
\label{BVgeneq}
\frac{1}{2}(W,W)=i\hbar\Delta W,
\end{eqnarray}
since the operator $V$ in (\ref{V}) vanishes on the surface $J_A=0$.

The generating functional of Green's functions (\ref{GSQ}), corresponding to
the vacuum functional (\ref{GFmodBV}), is given by
\begin{eqnarray}
\label{GFmodBV1}
 Z(J,\phi^*)
 \!=\!\int\! d\phi'\,d\phi^{*\prime}d\lambda'\exp\bigg\{\frac{i}{\hbar}
 \bigg[W(\phi',\phi^{*\prime},\lambda')\!+\!X(\phi',\phi^{*\prime},\lambda')
 \!+\!(\phi^{*\prime}-\phi^*)\lambda'\!+\!J\phi'\bigg]\!\bigg\},
\end{eqnarray}
where the variables $J_A$ acquire the meaning of sources to the fields $\phi^A$.
The functional $Z=Z(J,\phi^*)$ satisfies the Ward identities (\ref{Ward}), having
the component representation
\begin{eqnarray}
\label{VZ}
 J_A\frac{\delta Z}{\delta\phi^*_A}=\frac{i}{\hbar}J_A
 \left<\frac{\delta X}{\delta\phi^{*'}_A}\right>Z.
\end{eqnarray}

Taking into account the antibracket and delta-operator (\ref{deltabv}), entering
equation (\ref{BVgeneq}), one can restrict the consideration to functionals
$W$ independent of $\lambda^A$, thus supposing $W=W(\phi,\phi^*)$. Then the vacuum
functional (\ref{GFmodBV}) simplifies to
\begin{eqnarray}
\label{GFmodBV2}
 Z=\int
 d\phi\, d\phi^* d\lambda\exp\bigg\{\frac{i}{\hbar}
 \bigg[W(\phi,\phi^{*})+X(\phi,\phi^{*},\lambda)+\phi^{*}_A\lambda^A\bigg]\bigg\}.
\end{eqnarray}

There is a close connection between the above path integral
and the vacuum functional \cite{BBD} proposed to modify the gauge-fixing
procedure within the BV formalism.
In \cite{BBD}, it was shown that if the quantum action $W$ and the
action of gauge-fixing $X$ in the vacuum functional
\begin{eqnarray}
\label{GFBBD}
 Z=\int
 d\phi\;d\phi^* d\lambda\exp\bigg\{\frac{i}{\hbar}\bigg[W(\phi,\phi^*)
 +X(\phi,\phi^*,\lambda)\bigg]\bigg\}
\end{eqnarray}
satisfy the generating equation (\ref{BVgeneq}) of the BV formalism, then
(\ref{GFBBD}) does not depend of $X$. By virtue of (\ref{BVgeneq}) and the
component form (\ref{V}) of the operator $U$, one can see that the
functional $X'=X-\phi^*_A\lambda^A$ satisfies the generating equation
(\ref{GEqXSQ}), and therefore in terms of $X'$, the path integral
(\ref{GFmodBV2}) coincides with the vacuum functional (\ref{GFBBD}). This
implies that the quantization scheme of Section 2 can be considered as a
superfield form of the prescription (\ref{GFBBD}) proposed in \cite{BBD}.

Let us demonstrate that the vacuum functional (\ref{GFmodBV2}), equivalent
to (\ref{GFBBD}), provides an extension of the BV formalism in terms of
gauge-fixing. Indeed, the quantum action $W=W(\phi,\phi^*)$ satisfies the
generating equations (\ref{BVgeneq}) of the BV formalism. At the same time,
as we have mentioned above, the gauge functional $X(\Phi)=U\Psi(\Phi)$
satisfies (\ref{GEqXSQ}). Owing to the component form (\ref{V}) of the
operator $U$, with the functional $\Psi$ chosen to be independent of the
variables $\lambda^A$, the functional
\begin{eqnarray}
\nonumber
 X(\phi,\lambda)=-\frac{\delta\Psi(\phi)}{\delta \phi^A}\lambda^A
\end{eqnarray}
coincides with the action of gauge-fixing applied by the BV formalism in
the case of gauge conditions depending on the fields $\phi^A$, whereas the
variables $\lambda^A$ acquire the meaning of Lagrange multipliers
introducing the gauge. In the particular case of the gauge-fixing
functional $X=X(\phi,\lambda)$, the Ward identities (\ref{VZ}) for the
corresponding generating functional of Green's functions $Z=Z(J,\phi^*)$,
given by (\ref{GFmodBV1}), are reduced to
\begin{eqnarray}
\label{Wardbv}
 J_A\frac{\delta Z}{\delta\phi^*_A}=0.
\end{eqnarray}

As is well-known, the form of the Ward identities (\ref{Wardbv})
in the BV approach is not sensitive to the choice of gauge-fixing
and its dependence on the field-antifield variables
$(\phi^A,\phi^*_A)$. By comparison of (\ref{VZ}) and (\ref{Wardbv}),
we can see that fixing the gauge by equation (\ref{GEqXSQ}) actually
introduces an extension of the gauge-fixing procedure used in the
BV formalism.

The approach of Section 3 provides a generalization of the superfield
BRST-antiBRST quantization \cite{L} in the part of gauge-fixing. Indeed,
the operators $U^a$, $V^a$, $\Delta^a$ and the extended antibracket
$(\,\,,\,)^a$, given by (\ref{U&V}), (\ref{DeltaaExSQ}), (\ref{ABExSQ}),
are identical with the corresponding ingredients of the formalism \cite{L}.
At the same time, any functional
\begin{eqnarray} \nonumber
X(\Phi)=\frac{1}{2}\varepsilon_{ab}U^aU^b F(\Phi),
\end{eqnarray}
parameterized by an arbitrary Boson $F=F(\Phi)$, satisfies the generating
equation (\ref{GEqX1ExSQ}). This solution coincides with the action of
gauge-fixing used in \cite{L}.

Let us analyse the component form of the superfield prescription given in
Section 3 with the purpose of establishing its relation to other
quantization schemes. The component form of the superfields $\Phi^A(\theta)$
and supersources $\bar\Phi_A(\theta)$ is given by
\[
 \Phi^A(\theta)=\phi^A+\pi^{Aa}\theta_a+\lambda^A\theta^2,\;\;
\bar{\Phi}_A(\theta)=\bar{\phi}_A-\theta^a\phi^*_{Aa}-\theta^2J_A,
\]
\[
 \varepsilon(\phi^A)=\varepsilon(\lambda^A)=
 \varepsilon(\bar{\phi}_A)=\varepsilon(J_A)=\varepsilon_A,\;\;
 \varepsilon(\pi^{Aa})=\varepsilon(\phi^*_{Aa})=\varepsilon_A+1.
\]
In the component notations, the extended antibrackets (\ref{ABExSQ}),
(\ref{ABExSQ2}) have the form
\begin{eqnarray}
\label{br1}
(F,G)^a_1&=&\frac{\delta F}{\delta\phi^A}\;\frac{\delta G}{\delta\phi^*_{Aa}}
+\varepsilon^{ab}\frac{\delta F}{\delta\pi^{Ab}}\frac{\delta
G}{\delta\bar{\phi}_A}
-(-1)^{(\varepsilon(F)+1)(\varepsilon(G)+1)}
(F\leftrightarrow G)\;,\\
\label{br2}
(F,G)^a_2&=&\frac{\delta F}{\delta\phi^A}\;\frac{\delta G}{\delta\phi^*_{Aa}}
-(-1)^{(\varepsilon(F)+1)(\varepsilon(G)+1)}
(F\leftrightarrow G)
\end{eqnarray}
and the corresponding delta-operators (\ref{DeltaaExSQ}), (\ref{DeltaaExSQ2})
are given by
\begin{eqnarray}
\label{d1}
\Delta^a_1&=&(-1)^{\varepsilon_A}\frac{\delta_{\it l}}{\delta\phi^A}\;
\frac{\delta}{\delta\phi^*_{Aa}}+(-1)^{\varepsilon_A+1}\varepsilon^{ab}
\frac{\delta_{\it l}}{\delta\pi^{Ab}}\;\frac{\delta}{\delta\bar{\phi}_A}\,,\\
\label{d2}
\Delta^a_2&=&(-1)^{\varepsilon_A}\frac{\delta_{\it l}}{\delta\phi^A}\;
\frac{\delta}{\delta\phi^*_{Aa}}\,.
\end{eqnarray}
The antibracket and delta-operator (\ref{br1}), (\ref{d1}) coincide with
the corresponding ingredients of the (modified) triplectic formalism
\cite{3pl,mod3pl}, while (\ref{br2}), (\ref{d2}) are identical with those
of the Sp(2) covariant approach \cite{BLT}. The operators $U^a$,
$V^a$ in (\ref{U&V}) have the representation
\begin{eqnarray}
\nonumber
U^a&=&(-1)^{\varepsilon_A}\varepsilon^{ab}\lambda^A\frac{\delta_{\it l}}
{\delta\pi^{Ab}}-(-1)^{\varepsilon_A}\pi^{Aa}\frac{\delta_{\it
l}}{\delta\phi^A}\,,\\
\label{V2}
V^a&=&\varepsilon^{ab}\phi^*_{Ab}\frac{\delta}{\delta\bar{\phi}_A}-
J_A\frac{\delta}{\delta\phi^*_{Aa}}\,.
\end{eqnarray}
Taking into account the component form of the integration measure in (\ref{ZExSQ})
\begin{eqnarray}
\nonumber
d\Phi\, d\bar{\Phi}\,\rho(\bar{\Phi})=d\phi\, d\phi^* d\pi\, d\bar{\phi}\,
d\lambda\, dJ\,\delta(J)
\end{eqnarray}
and the functional $\bar\Phi\Phi$ in (\ref{BilFExSQ})
\begin{eqnarray}
\nonumber
\bar{\Phi}\Phi=\bar{\phi}_A\lambda^A+\phi^*_{Aa}\pi^{Aa}-J_A\phi^A,
\end{eqnarray}
we represent the vacuum functional (\ref{GEqExSQ}) as the following
path integral:
\begin{eqnarray}
\label{vacf}
 \!\!\!Z\!=\!\int\!\!
 d\phi\, d\phi^* d\bar{\phi}\, d\pi d\lambda
 \exp\!\bigg\{\frac{i}{\hbar}\bigg[W(\phi,\phi^*,\bar{\phi},\pi,\lambda)
 \!+\! X(\phi,\phi^*,\bar{\phi},\pi,\lambda)
 \!+\!\phi^*_{Aa}\pi^{Aa}
 \!+\!\bar\phi_A\lambda^A\bigg]\!\bigg\},
\end{eqnarray}
where integration over $J_A$ has been carried out, thus placing the
functionals $W$, $X$ on the surface $J_A=0$. The functionals
$W=W(\phi,\phi^{*},\bar\phi,\pi,\lambda)$,
$X=X(\phi,\phi^{*},\bar\phi,\pi,\lambda)$ satisfy the generating equations
(\ref{GEq1ExSQ}), (\ref{GEqX1ExSQ}), restricted to $J_A=0$, where it has
been assumed that the corresponding antibracket and delta-operator are
chosen in the form (\ref{br1}), (\ref{d1}), which guarantees the
gauge-independence of the S-matrix. The restriction $J_A=0$ does not
affect the form of (\ref{GEqX1ExSQ}), whereas (\ref{GEq1ExSQ}) is
transformed to a similar equation with the operator $V^a$ (\ref{V2})
replaced by a truncated one,
\begin{eqnarray}
\label{vtrunc}
V^a=\varepsilon^{ab}\phi^*_{Ab}\frac{\delta}{\delta\bar{\phi}_A}.
\end{eqnarray}

The generating functional of Green's functions (\ref{GSQ1}), corresponding to
the vacuum functional (\ref{vacf}), is given by
\begin{eqnarray}
\label{gf}
 \!\!Z(J,\phi^*,\bar\phi)\!=\!\!\int\!
 d\phi'\,d\phi^{*\prime}d\bar{\phi}'\,d\pi' d\lambda'
 \exp\!\bigg\{\frac{i}{\hbar}\bigg[W\!+\!X\!+\!
 (\phi^{*\prime}-\phi^{*})\pi^{\prime}
 \!+\!(\bar\phi'-\bar\phi)\lambda^{\prime}\!+\!
 J\phi^{\prime}\bigg]\!\bigg\},
\end{eqnarray}
 where $W$, $X$ depend on the integration variables.
 The functional $Z=Z(J,\phi^*,\bar\phi)$ satisfies the Ward identities
 (\ref{Ward1}), having the component form
\begin{eqnarray}
\label{VZ1}
 J_A\frac{\delta Z}{\delta \phi^{*}_{Aa}}-\varepsilon^{ab}\phi^{*}_{Ab}
 \frac{\delta Z}{\delta\bar\phi_A}=\frac{i}{\hbar}J_A
 \left<\frac{\delta X}{\delta\phi^{*\prime}_{Aa}}\right>Z
 -\frac{i}{\hbar}\varepsilon^{ab}\phi^{*}_{Ab}
 \left<\frac{\delta X}{\delta\bar\phi^{\prime}_A}\right>Z.
\end{eqnarray}

Taking into account the antibracket (\ref{br1}), the delta-operator
(\ref{d1}), and the truncated operator $V^a$ (\ref{vtrunc}), entering
the equation that determines the quantum action in
(\ref{vacf}), one can restrict the consideration to functionals $W$
independent of $\lambda^A$, thus supposing $W=W(\phi,\phi^*,\bar\phi,\pi)$.
Then the vacuum functional (\ref{vacf}) simplifies to
\begin{eqnarray}
\label{vacf1}
 \!\!\!Z\!=\!\int\!\!
 d\phi\, d\phi^* d\bar{\phi}\, d\pi d\lambda
 \exp\bigg\{\frac{i}{\hbar}\bigg[W(\phi,\phi^*,\bar{\phi},\pi)
 \!+\! X(\phi,\phi^*,\bar{\phi},\pi,\lambda)
 \!+\!\phi^*_{Aa}\pi^{Aa}
 \!+\!\bar\phi_A\lambda^A\bigg]\!\bigg\},
\end{eqnarray}

There is a close connection between the above path integral and the vacuum
functional of the modified triplectic approach \cite{mod3pl}
\begin{eqnarray}
\label{tripl}
 Z=\int
 d\phi\;d\phi^* d\pi\, d\bar{\phi}\;d\lambda
 \exp\bigg\{\frac{i}{\hbar}\bigg[W(\phi,\phi^*,\bar{\phi},\pi)
 + X(\phi,\phi^*,\bar{\phi},\pi,\lambda) + \phi^*_{Aa}
 \pi^{Aa}\bigg]\bigg\},
\end{eqnarray}
proposed to extend the gauge-fixing procedure of the Sp(2) covariant
formalism \cite{BLT}. In (\ref{tripl}), $W$ obeys the same equation that
is satisfied by the quantum action in (\ref{vacf1}), while $X$ is a
gauge-fixing functional satisfying (\ref{GEqX1ExSQ}), with $U^a$ in
(\ref{V2}) replaced by the truncated operator (corresponding to $\lambda^A=0$)
\begin{eqnarray}
\label{UaMTQ}
U^a=-(-1)^{\varepsilon_A}\pi^{Aa}\frac{\delta_{\it l}}{\delta\phi^A}.
\end{eqnarray}
In \cite{mod3pl}, it was demonstrated that the vacuum functional (\ref{tripl}) does
not depend on the choice of $X$. It is easy to see that the functional
$X' = X - \bar{\phi}_A\lambda^A$ obeys the generating equation
(\ref{GEqX1ExSQ}), with the complete operator $U^a$ (\ref{V2}), and
therefore in terms of $X'$, the path integral (\ref{vacf1}) is identical
with the vacuum functional (\ref{tripl}). This means that the quantization
scheme of Section 3 can be considered as a superfield form of the prescription
(\ref{tripl}) proposed in \cite{mod3pl}

Let us show that the vacuum functional (\ref{vacf1}), equivalent to
(\ref{tripl}), provides an extension of the gauge-fixing procedure used in
the Sp(2) covariant formalism \cite{BLT}. For this purpose, we restrict the
consideration to the subclass of functionals $W$ in (\ref{vacf1}) chosen to
be independent of the variables $\pi^{Aa}$,
\begin{eqnarray}
\label{restr}
 \frac{\delta W}{\delta\pi^{Aa}}=0,
\end{eqnarray}
thus supposing $W=W(\phi,\phi^*,\bar\phi)$, where the set of variables
$(\phi^A,\phi^*_{Aa},\bar\phi_A)$ is identical with the field-antifield
variables of the Sp(2) covariant formalism.
The restriction (\ref{restr}) simplifies the
corresponding equation for $W$. Namely, in (\ref{GEq1ExSQ}) one replaces
the antibracket and delta-operator by (\ref{br2}), (\ref{d2}), respectively.
Taking into account the truncated operator $V^a$ (\ref{vtrunc}), substituted
into (\ref{GEq1ExSQ}) in the given case, we find that the equation for
$W=W(\phi,\phi^*,\bar\phi)$ coincides with the equation
that determines the quantum action in the Sp(2) covariant formalism. The vacuum
functional (\ref{vacf1}) simplifies to
\begin{eqnarray}
\label{Sp(2)}
 \!\!\!Z\!=\!\int\!\!
 d\phi\, d\phi^* d\bar{\phi}\, d\pi d\lambda
 \exp\bigg\{\frac{i}{\hbar}\bigg[W(\phi,\phi^*,\bar{\phi})
 \!+\! X(\phi,\phi^*,\bar{\phi},\pi,\lambda)
 \!+\!\phi^*_{Aa}\pi^{Aa}
 \!+\!\bar\phi_A\lambda^A\bigg]\!\bigg\}.
\end{eqnarray}
As mentioned above, the gauge functional
$X(\Phi)=\frac{1}{2}\varepsilon_{ab}U^aU^b F(\Phi)$ satisfies
(\ref{GEqX1ExSQ}). Owing to the component form (\ref{V2}) of the operator
$U^a$, with the functional $F$ chosen to be independent of the variables
$(\pi^{Aa},\lambda^A)$, the functional
\begin{eqnarray*}
 X(\phi,\pi,\lambda)=-\frac{\delta F}{\delta\phi^A}\lambda^A
 -\frac{1}{2}\varepsilon_{ab}\pi^{Aa}{\frac {\delta^2 F}{\delta\phi^A
 \delta\phi^B}}\pi^{Bb}
\end{eqnarray*}
coincides with the action of gauge-fixing applied by the Sp(2)
covariant formalism in the case of gauge conditions depending on the fields
$\phi^A$, where $(\pi^{Aa},\lambda^A)$ acquire the meaning of auxiliary
gauge-fixing variables. In the specified case of the gauge-fixing
functional $X=X(\phi,\pi,\lambda)$, the Ward identities (\ref{VZ1}) for the
corresponding generating functional of Green's functions
$Z=Z(J,\phi^*,\bar\phi)$, given by (\ref{gf}), are reduced to
\begin{eqnarray}
\label{Wardbvx}
 J_A\frac{\delta Z}{\delta \phi^{*}_{Aa}}-\varepsilon^{ab}\phi^{*}_{Ab}
 \frac{\delta Z}{\delta\bar\phi_A}=0.
\end{eqnarray}

Notice that the form of the Ward identities (\ref{Wardbvx})
in the Sp(2) covariant approach is not sensitive to the choice
of gauge-fixing and its dependence on the field-antifield variables
$(\phi^A,\phi^*_{Aa},\bar\phi_A)$. By comparison of (\ref{VZ1}) and
(\ref{Wardbvx}), we can see that fixing the gauge by equation (\ref{GEqX1ExSQ})
actually introduces an extension of the gauge-fixing procedure used in the
Sp(2) covariant approach.

\paragraph{Acknowledgement}
 The work was partially supported by the Russian
 Ministry of Education (Fundamental Sciences Grant E00-3.3-461).
 The work of P.M.L. was also supported by INTAS, grant 99-0590.
 P.Yu.M. is grateful to FAPESP, grant 02/00423-4.

\end{document}